\documentclass[prb, preprint,unsortedaddress]{revtex4}

\usepackage{amsfonts, amsbsy, amssymb, amsmath, graphicx, float} 

\usepackage{color}
\usepackage{float}
\usepackage{amsthm}

\usepackage{pifont}      
\reversemarginpar

\def \cK {\mathcal{K}}

\begin{document}


\title{Roaming at Constant Kinetic Energy: Chesnavich's Model and the Hamiltonian Isokinetic  Thermostat}



\author{Vladim{\'i}r Kraj{\v{n}}{\'a}k}
\email[]{v.krajnak@bristol.ac.uk}
\affiliation{School of Mathematics \\
University of Bristol\\Bristol BS8 1TW\\United Kingdom}

\author{Gregory S. Ezra}
\email[]{gse1@cornell.edu}
\affiliation{Department of Chemistry and Chemical Biology \\
Baker Laboratory\\Cornell University\\ Ithaca, NY 14853\\USA}

\author{Stephen Wiggins}
\email[]{s.wiggins@bristol.ac.uk}
\affiliation{School of Mathematics \\
University of Bristol\\Bristol BS8 1TW\\United Kingdom}



\begin{abstract}

We consider the roaming mechanism for chemical reactions under the nonholonomic constraint of constant kinetic energy. Our study is carried out in the context of the Hamiltonian isokinetic thermostat applied to Chesnavich's model for an ion-molecule reaction. Through an analysis of phase space structures we show that imposing the nonholonomic constraint does not prevent the system from exhibiting roaming dynamics, and that the origin of the roaming mechanism turns out to be analogous to that found in the previously studied Hamiltonian case.

\end{abstract}


\maketitle


\section{Introduction}
\label{sec:intro}

The roaming mechanism for chemical reactions was introduced in reference to the description of the photodissociation of formaldehyde \cite{zee:1664, townsend2004roaming, bowman}. After excitation of the formaldehyde molecule by a laser pulse, a single CH bond begins to stretch. Rather than proceed directly to dissociation, the hydrogen atom rotates around the molecular fragment in a region of the potential energy surface where it is subject to weak forces.  The corresponding motion of the hydrogen atom was termed `roaming'. At a later stage the roaming hydrogen atom encounters the bound hydrogen atom and undergoes an  abstraction reaction. The resulting H$_2$ molecule then separates from the CO fragment. This reaction is then said to occur by the roaming mechanism.

Since the pioneering formaldehyde studies a similar roaming mechanism has been observed to occur in a number of chemical reactions. Many of these reactions are described in several review articles\cite{suits2008, bowman2011roaming,Bowman2011Suits, BowmanRoaming, mauguiere2017roaming}.

Essentially all of these roaming reactions have been studied at constant total energy. 
The goal of this article is to study analogous roaming reaction mechanisms at constant kinetic energy.
Specifically, we study Chesnavich's model for an ion-molecule reaction subjected to a Hamiltonian isokinetic thermostat.

Chesnavich's empirical model for the ion-molecule reaction CH$_4^+$ $\rightleftharpoons$ CH$_3^+$ $+$ H was 
introduced in reference \onlinecite{Chesnavich1986}. 
A detailed derivation of this model can be found in reference \onlinecite{ezra2019chesnavich}. 
Chesnavich's model captures many of the essential aspects of the roaming mechanism. It describes the situation where a hydrogen atom separates from the rigid CH$_3^+$ core and, instead of dissociating, roams in a region of nearly constant potential only to return to the core.  While Chesnavich's model does not accurately describe the intramolecular abstraction and subsequent dissociation, it has nevertheless provided significant insight into the roaming process, see, for example, 
refernces \onlinecite{mauguiere2014multiple, mauguiere2014roaming, krajnak2018phase, krajnak2018influence}.

In the present work we apply a Hamiltonian version of the isokinetic thermostat to Chesnavich's model. 
This thermostat is not as widely used as the Nos\'{e}-Hoover thermostat, and its many variants, 
but the non-Hamiltonian version of the isokinetic thermostat has been developed and applied to a 
variety of problems of interest in chemistry by Minary et al.\ \cite{Minary03a,Minary03b}. 
In this thermostat, the particle momenta are subject to a nonholonomic constraint that 
keeps the kinetic energy constant.  The resulting dynamics (if ergodic) generates 
a canonical distribution in configuration space, where the associated temperature is 
related to the value of the conserved kinetic energy \cite{Morriss98}. 
A Hamiltonian version of the isokinetic thermostat was given by Dettmann \cite{Dettmann96,Morriss98} 
(see also \cite{Litniewski93,Morishita03}).  

The Hamiltonian isokinetic thermostat dynamics explored in our 
paper is relatively unfamiliar, although, as noted, there has 
been some relevant earlier work in the chemical literature \cite{Minary03a,Minary03b}.
The basic theory \cite{Dettmann96,Morriss98}  shows that Hamiltonian dynamics 
in terms of a set of variables related to the physical variables by
a \emph{noncanonical} tranformation, together with the
use of the exponentiated potential results in a canonical distribution
in the physical configuration variables (assuming, as usual, ergodic dynamics).
The unfamiliar aspect of the dynamics lies in the imposition of a nonholonomic 
constraint on the original variables; namely, constant kinetic
energy.
Again, the general theory shows that the effective temperature is proprtional to
the value of the conserved kinetic energy.

It is in this sense that the nonholonomic dynamics explored here for
the
Chesnavich model is effectively at constant temperature.
We do not claim that this dynamics is the same as obtained using a Boltzmann distribution 
for the physical momenta; nevertheless, one point of our paper is that the phase space structure of
the nonholonomic system can in this case be analyzed using methods
employed for the microcanonical case, and an analogous roaming mechanism explored.

For our purposes the Hamiltonian formulation of the isokinetic thermostat then has several advantages. 
There has been considerable development in our understanding of phase space structures 
governing chemical reaction dynamics in recent years, see, for example, 
references \onlinecite{ wwju, uzer2002geometry, WaalkensSchubertWiggins08, wiggins2016}.
This phase space structure approach has been applied to the Hamiltonian isokinetic thermostat 
formalism  \cite{ezra2008impenetrable, collins2010phase}. In this paper we continue 
these developments in our study of the Hamiltonian phase space structure associated with the 
roaming mechanism as exhibited in Chesnavich's model subjected to a Hamiltonian isokinetic thermostat.

This paper is outlined as follows:  In Sec.\ \ref{sec: HIK} we discuss the Hamiltonian formulation of the isokinetic thermostat for a general system of the form `kinetic $+$ potential'.   In Sec.\ \ref{sec:CMR} we introduce Chesnavich's model for roaming in CH$_4^{+}$, and discuss the dynamical origin of roaming in terms of families of unstable periodic orbits and their associated invariant manifolds.  In  Sec.\ \ref{sec:ICM}  we introduce the Hamiltonian for Chesnavich's model subject to an isokinetic thermostat.  In the absence of self-retracing orbits for the thermostatted system, we find two families of periodic orbits that are relevant to the roaming phenomenon.  Analysis of escape times and Lagrangian descriptors  shows that roaming does indeed occur in the thermostatted system, via a mechanism analogous to that found in the Hamiltonian case.  Sec.\ \ref{sec:conc} concludes.

\section{The Hamiltonian Isokinetic Thermostat}
\label{sec: HIK}

\subsection{Thermostatted Hamiltonian}
We consider a Hamiltonian system
\begin{equation}
H(q,p) = T(q,p) + \Phi(q),
\label{eq:Ham}
\end{equation}
where the  position-dependent kinetic energy $T(q,p)$ is a quadratic form of the momenta and $\Phi(q)$ is a potential energy. The aim of the isokinetic thermostat is to constrain $T$ to a constant value $T_0^2>0$.

We define the isokinetic Hamiltonian $\cK$ by
\begin{equation}
 \cK(q,\pi) = e^{\Phi}T(q,\pi) - e^{-\Phi}T_0^2.
\end{equation}
Via the non-canonical transformation,
\begin{equation}
 \pi=e^{-\Phi}p
\label{eq:transf}
\end{equation}
the level set $\cK=0$ corresponds to constant kinetic energy $T=T_0^2$ in the system \eqref{eq:Ham}. Note that since $T$ is a quadratic form of the momenta, from \eqref{eq:transf} we have
\begin{equation}
 T(q,\pi)=T(q,e^{-\Phi}p)=e^{-2\Phi}T(q,p),
\end{equation}
and hence
\begin{equation}
 e^{\Phi}T(q,\pi) - e^{-\Phi}T_0^2=e^{-\Phi}\left(T(q,p) - T_0^2\right).
\end{equation}

Due to the Hamiltonian structure of the system, $\dot{\cK}=0$ along the solutions of Hamilton's equations
\begin{equation}
 \begin{split}
 \dot{q}&=\frac{\partial \cK}{\partial \pi}=e^{\Phi}\frac{\partial T(q,\pi)}{\partial \pi},\\
 \dot{\pi}&=-\frac{\partial \cK}{\partial q}= -e^{\Phi}\frac{\partial \Phi}{\partial q}T(q,\pi) -e^{\Phi}\frac{\partial T(q,\pi)}{\partial q} -e^{-\Phi}\frac{\partial \Phi}{\partial q}T_0^2,\\
 \end{split}
 \label{eq:eqKham}
\end{equation}

\subsection{Time scaling}
\label{sec:scaling}
 In this section, we show that the dynamics defined by equations \eqref{eq:eqKham} produce equivalent dynamics regardless of the 
 value of $T_0^2$. We prove that a suitable scaling of time and momenta transforms \eqref{eq:eqKham} into equations with $T_0=1$.
 
 Consider the scaling of time and momenta by
\begin{equation}
 s=T_0t, \quad \Pi=\frac{\pi}{T_0}.
 \label{eq:tpscale}
\end{equation}
 From \eqref{eq:tpscale} and
\begin{equation}
 ds=T_0dt,
 \label{eq:timescale}
\end{equation}
 it follows that
 \begin{equation}
  \begin{split}
 \frac{dq}{ds}&=\frac{dt}{ds}\frac{dq}{dt}=\frac{1}{T_0}\dot{q},\\
 \frac{d\Pi}{ds}&=\frac{dt}{ds}\frac{d\Pi}{d\pi}\frac{d\pi}{dt}=\frac{1}{T^2_0}\dot{\pi}.  
 \end{split}
 \label{eq:derivs}
\end{equation}
 
Using \eqref{eq:eqKham}, \eqref{eq:derivs} and
\begin{equation}
 T(q,\Pi)=\frac{1}{T^2_0}T(q,\pi),
 \label{eq:Pitopi}
\end{equation}
 we can write the equations of motion in $s$ and $\Pi$ as
\begin{equation}
\begin{split}
 \frac{dq}{ds}&=\frac{1}{T_0}e^{\Phi}\frac{\partial T(q,\pi)}{\partial \pi}={T_0}e^{\Phi}\frac{\partial T(q,\Pi)}{\partial \pi}=e^{\Phi}\frac{\partial T(q,\Pi)}{\partial \Pi},\\
 \frac{d\Pi}{ds}&=-e^{\Phi}\frac{\partial \Phi}{\partial q}\frac{1}{T^2_0}T(q,\pi) -e^{\Phi}\frac{1}{T^2_0}\frac{\partial T(q,\pi)}{\partial q} -e^{-\Phi}\frac{\partial \Phi}{\partial q}\frac{1}{T^2_0}T_0^2,\\
 &= -e^{\Phi}\frac{\partial \Phi}{\partial q}T(q,\Pi) -e^{\Phi}\frac{\partial T(q,\Pi)}{\partial q} -e^{-\Phi}\frac{\partial \Phi}{\partial q}.\\
 \end{split}
\end{equation}

These equations correspond to the isokinetic Hamiltonian with $T_0=1$
\begin{equation}
 \cK_1(q,\Pi) = e^{\Phi}T(q,\Pi) - e^{-\Phi},
\end{equation}
which is related to $\cK$ via
\begin{equation}
 \cK_1(q,\Pi) = \frac{1}{T^2_0}\cK(q,\pi).
\end{equation}
Clearly the dynamics of $\cK$ and $\cK_1$ is equivalent and only differs by scaling by a constant factor. 

\section{Chesnavich's Model and Roaming}
\label{sec:CMR}
In this section we introduce Chesnavich's model for the ion-molecule reaction CH$_4^+$ $\rightleftharpoons$ CH$_3^+$ $+$ H and recall known results about roaming in this system. The model was introduced by Chesnavich  \cite{Chesnavich1986} to investigate the transition from vibration/librational motion of the H-atom  in a deep potential well representing CH$_4^+$ to nearly-free rotation in a flat and rotationally symmetric region representing the dissociated CH$_3^++$H.

\subsection{Chesnavich's Model Hamiltonian}
\label{sec:CMH}
Chesnavich's CH$_4^+$ model is a Hamiltonian system with 2 degrees of freedom, consisting of a rigid CH$_3^+$ molecule and a mobile H atom. The system is defined by the Hamiltonian
\begin{equation}
H(r,\theta,p_r, p_\theta) = \frac{1}{2} \frac{p_r^2}{\mu} + \frac{1}{2}p_\theta^2 \left(\frac{1}{\mu r^2}+\frac{1}{I_{CH_3}}\right) + U(r,\theta),
\label{eq:chesHam}
\end{equation}
where $(r, \theta, \phi)$ are polar coordinates describing the position of the H-atom in a body-fixed frame attached to the CH$_3^+$ core
(coordinate $\phi$ is ignorable in this model).  The reduced mass of the system $\mu=\frac{m_{CH_3}m_{H}}{m_{CH_3}+m_{H}}$, where $m_{H}=1.007825$ u and $m_{CH_3}=3m_{H}+12.0$ u, and the moment of inertia of the rigid body CH$_3^+$ $I_{CH_3}=2.373409$~u\AA$^2$.

\begin{figure}
 \includegraphics[width=.5\textwidth]{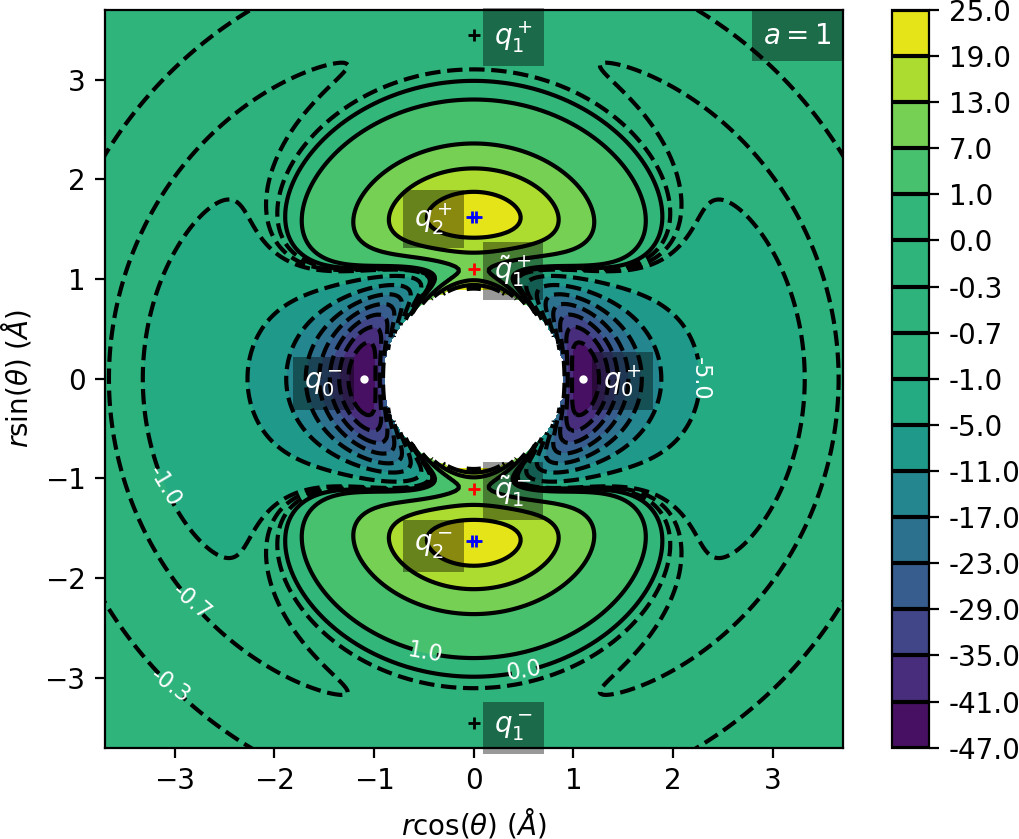}
 \caption{Contour plot of Chesnavich's potential energy surface $U$ for $a=1$. Dashed lines correspond to $U<0$, solid lines correspond to $U\geq0$. Contours correspond to values of potential shown on the colorbar right, with some values indicated in the plot.
 }
 \label{fig:pot}
\end{figure}

The potential $U(r,\theta)$ consists of a radial long range potential $U_{CH}$ and a short range potential $U_{coup}$ that models the short range anisotropy of the rigid CH$_3^+$ body.
\begin{equation}\label{eq:U}
 U(r,\theta ) = U_{CH} (r) + U_{coup} (r,\theta).
\end{equation}
It is characterised by two deep wells corresponding to bound CH$_4^+$, two areas of high potential and a flat area beyond them as shown in Fig. \ref{fig:pot}.

The long range potential is
\begin{equation}\label{eq:UCH}
 U_{CH} (r) =  \frac{D_e}{c_1 - 6} \left( 2 (3-c_2) e^{c_1 (1-x)}  - \left( 4 c_2 - c_1 c_2 + c_1 \right) x^{-6} - (c_1 - 6) c_2 x^{-4} \right), 
\end{equation}
with parameter values as used in the original work \cite{Chesnavich1986}:$x = \frac{r}{r_e}$, dissociation energy $D_e=47$ kcal/mol, equilibrium bond length $r_e=1.1$ \AA, $c_1=7.37$ and $c_2=1.61$.
The short range hindered rotor  potential $U_{coup}$ is defined by 
\begin{equation}\label{eq:Ucoup}
 U_{coup} (r,\theta) = \frac{U_e e^{-a(r-r_e)^2}}{2} (1 - \cos 2 \theta ),
\end{equation}
where $U_e=55$ kcal/mol is the equilibrium barrier height. The distance at which transition occurs from rotation to vibration is determined by the parameter $a$ (in \AA$^{-2}$). Various values of $a$ have been considered in previous works, namely $a=1$ \cite{Chesnavich1986,mauguiere2014roaming,mauguiere2014multiple,krajnak2018phase}, $a=4$ \cite{Chesnavich1986,mauguiere2014roaming} and a range of values $0.7\leq a\leq 8$. \cite{krajnak2018influence}

The CH$_3^+$ is a symmetric top in Chesnavich's model. Although the range of coordinate $\theta$ is $0 \leq \theta \leq \pi$, in the planar (zero overall angular momentum) version of the model the range of $\theta$ is extended to $0 \leq \theta \leq 2 \pi$, and the potential has four-fold symmetry
\begin{equation}
U(r,\theta)=U(r,-\theta)=U(r,\pi-\theta)=U(r,\pi+\theta).
\label{eq:sym}
\end{equation}

The potential admits four pairs of equilibrium points pairwise related by symmetry \eqref{eq:sym}, as listed in Tab. \ref{table:equil} and shown in Fig. \ref{fig:pot}.
  
  \begin{table}
    \begin{center}
    \begin{tabular}{c|c|c|c|c}
    Energy (kcal mol$^{-1}$) & $r$ (\AA) & $\theta$ (radians) & Significance & Label \\
    \hline
    $-47$ & $1.1$ & $0$ & potential well & $q_0^+$ \\
    $-0.63$ & $3.45$ & $\pi/2$ & isomerisation saddle & $q_1^+$ \\
    $8$ & $1.1$ & $\pi/2$ & isomerisation saddle & $\widetilde{q}_1^+$ \\
    $22.27$ & $1.63$ & $\pi/2$ & local maximum & $q_2^+$ \\
    \end{tabular}
    \end{center}
    \caption{Equilibrium points of the potential $U(r, \theta)$.}
    \label{table:equil}
   \end{table}

\subsection{Roaming in Chesnavich's Model}
\label{sec:roamingChes}
As mentioned in Section \ref{sec:intro}, we are interested in the roaming mechanism wherein the hydrogen 
atom separates from the CH$_3^+$ core only to return to the vicinity of the core before dissociating. Here we review the dynamical definition of roaming as introduced in \cite{mauguiere2014roaming}, which is based on periodic orbits as invariant structures having dynamical relevance.

\begin{figure}
 \includegraphics[width=.5\textwidth]{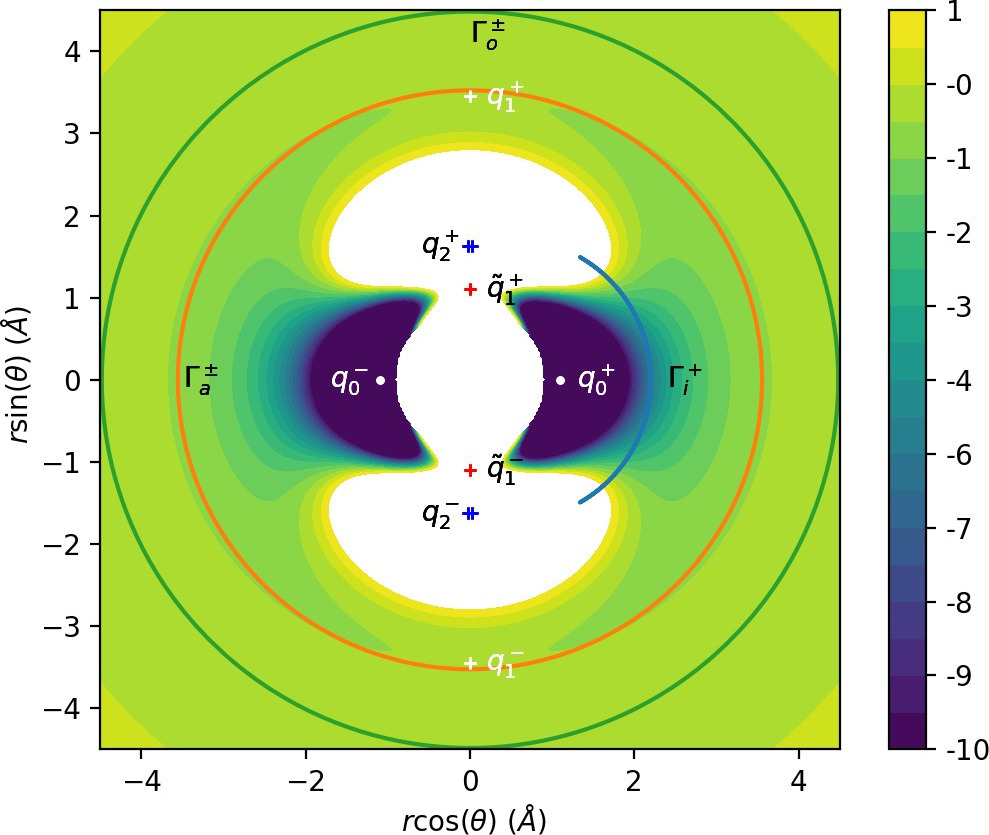}
 \caption{Configuration space projections of inner ($\Gamma^i$), middle ($\Gamma^a$) and outer ($\Gamma^o$) periodic orbits for $E=5$.}
 \label{fig:orbitse5}
\end{figure}

In the relevant energy interval $0\leq E\leq 5$, there are three important families of periodic orbits \cite{mauguiere2014multiple}. At any given fixed energy $0\leq E\leq 5$, there are three pairs of important periodic orbits present in this system, pairwise related by symmetry \eqref{eq:sym}, as shown in Fig. \ref{fig:orbitse5}. Therefore we refer to a continuum of periodic orbits parametrised by energy as a family of periodic orbits. We will refer to these families as the inner ($\Gamma^i$), middle ($\Gamma^a$) and outer ($\Gamma^o$) periodic orbits. It is important to note that none of the orbits is directly related to a saddle point of the system.

Their significance is as follows:
\begin{itemize}
 \item $\Gamma^i$: Delimits the potential wells that correspond to CH$_4^+$ isomers. The orbits oscillate around the axes $\theta=0$ and $\theta=\pi$.
 \item $\Gamma^a$: Two rotational orbits with opposite orientation - one clockwise, one counter-clockwise that are crucial for defining roaming.
 \item $\Gamma^o$: Centrifugal barrier delimiting the region of dissociated states. Two rotational orbits with opposite orientation - one clockwise, one counter-clockwise.
\end{itemize}
Roaming was shown \cite{krajnak2018phase} to disappear for $E\geq 2.5$. In the energy interval $0< E< 2.5$, all of the above-mentioned periodic orbits are unstable. Define the dividing surfaces using DS$^i$, DS$^a$ and DS$^o$ as the set of all phase space points $(r,\theta,p_r, p_\theta)$ that satisfy
$$H(r,\theta,p_r, p_\theta)=E,$$
and their configuration space projections $(r,\theta)$ coincide with the configuration space projections of $\Gamma^i$, $\Gamma^a$ and $\Gamma^o$ respectively. Due to the instability of the orbits, the resulting dividing surfaces do not admit local recrossing.

A roaming trajectory is then defined as a trajectory that crosses DS$^a$ an odd number of times between leaving the potential well and dissociating. Trajectories that return to their region of origin cross DS$^a$ an even number of times - isomerising trajectories returning to either of the potential wells, nonreactive trajectories return to dissociated states.

As explained in reference \onlinecite{krajnak2018phase}, the oscillatory nature of 
$\Gamma^i$ implies that DS$^i$ consists of two spheres, while the rotational nature of $\Gamma^a$ and $\Gamma^o$ means that DS$^a$ and DS$^o$ are tori\cite{mauguiere2016ozone}. Each sphere can be divided using the corresponding periodic orbit into two hemispheres and each torus can be divided using both corresponding periodic orbits into two annuli\cite{mauguiere2016ozone}. All hemispheres and annuli are surfaces of unidirectional flux, for example all trajectories leaving the potential well cross the same (outward) hemisphere of DS$^i$, while all trajctories entering the potential well cross the other (inward) hemisphere of DS$^i$.

Roaming can hereby be reformulated as a transport problem in phase space. Every trajectory leaving the well must cross the outward hemisphere of DS$^i$ and every trajectory that dissociates must cross the outward annulus of DS$^o$. Dissociation of a CH$_4^+$ molecule is therefore equivalent to the transport of trajectories from the outward hemisphere of DS$^i$ to the outward annulus of DS$^o$. Roaming involves crossing the inward annulus of DS$^a$, because between two crossings of the outward annulus trajectories must cross the inward annulus and vice versa.

\begin{figure}
 \includegraphics{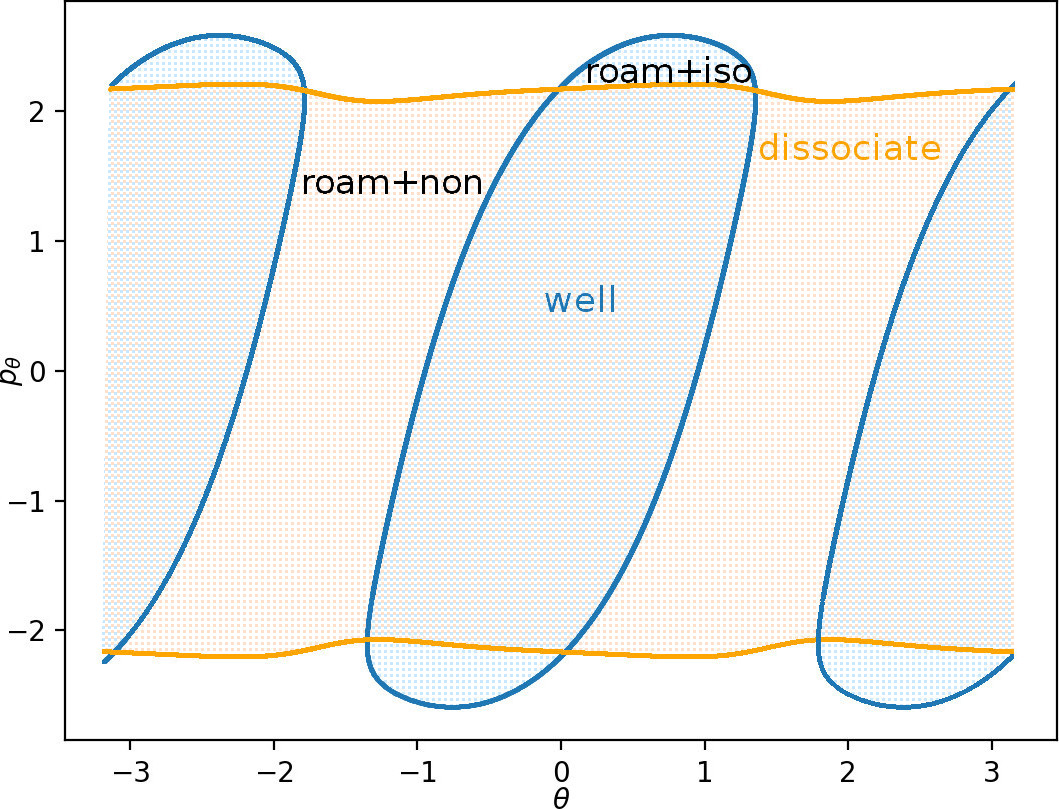}
 \caption{Intersection of invariant manifolds of $\Gamma^i$ (blue) and $\Gamma^o$ (orange) with the outward annulus of DS$^a$ for $E=1$. Trajectories that just left the potential well are shown in blue, immediately dissociating trajectories in orange. Roaming and isomerising trajectories in the blue area just left the well and do not dissociate immediately, while roaming and dissociating trajectories in the orange are dissociate immediately but did not just leave the potential wells.}
 \label{fig:DSa}
\end{figure}

Transport of trajectories in the neighbourhood of an unstable periodic orbit (or NHIM in general) is governed by invariant manifolds of this orbit \cite{wwju, uzer2002geometry, WaalkensSchubertWiggins08, wiggins2016}. It was shown\cite{krajnak2018phase, krajnak2018influence} that the roaming phenomenon involves a heteroclinic intersection of the invariant manifolds of $\Gamma^i$ and $\Gamma^o$. The condition  $H(r,\theta=0,p_r>0, p_\theta=0)$ defines an invariant subsystem that consists of precisely one dissociating trajectory for every fixed $E>0$. Therefore if the invariant manifolds of $\Gamma^i$ and $\Gamma^o$ do not intersect, the former are contained in interior of the latter and each trajectory leaving the potential well dissociates immediately. An intersection assures that some trajectories leaving the well do not dissociate immediately and return to DS$^a$ as illustrated in Fig. \ref{fig:DSa}. This allows for roaming and isomerisation.

\section{Isokinetic Chesnavich Model}\label{sec:ICM}

\subsection{Hamiltonian and periodic orbits for thermostatted Chesnavich model}

The isokinetic Hamiltonian $\cK$ for Chesnavich's model is defined as follows: 
\begin{equation}
 \cK(r,\pi_r,\theta,\pi_\theta) = \frac{1}{2}e^{U}\left(\frac{\pi_r^2}{\mu}+\pi_\theta^2\left(\frac{1}{\mu r^2}+\frac{1}{I_{CH_3}}\right)\right) - \frac{1}{2}e^{-U},
\label{eq:chesKHam}
\end{equation}
where $U=U(r,\theta)$ is Chesnavich's potential energy and 
\begin{equation}
 \pi_r=e^{-U}p_r,\quad \pi_\theta=e^{-U}p_\theta.
 \label{eq:isotransf}
\end{equation}
The level set $\cK=0$ corresponds to the surface of constant kinetic energy
\begin{equation}
 \frac{1}{2}\frac{p_r^2}{\mu}+\frac{1}{2}p_\theta^2\left(\frac{1}{\mu r^2}+\frac{1}{I_{CH_3}}\right)=\frac{1}{2},
 \label{eq:constantT}
\end{equation}
in system \eqref{eq:chesHam}.
Equations of motion in the isokinetic system are
\begin{equation}
 \begin{split}
 \dot{r}&=\frac{\partial \cK}{\partial {\pi}_r}=e^{U}\frac{\pi_r}{\mu},\\
 \dot{\pi}_r&=-\frac{\partial \cK}{\partial r}= -\frac{1}{2}e^{U}\frac{\partial U}{\partial r}\left(\frac{\pi_r^2}{\mu}+\pi_\theta^2\left(\frac{1}{\mu r^2}+\frac{1}{I_{CH_3}}\right)\right)  + \frac{1}{2}e^{U} \frac{2}{\mu r^3}\pi_\theta^2 -\frac{1}{2}e^{-U}\frac{\partial U}{\partial r},\\
 \dot{\theta}&=\frac{\partial \cK}{\partial {\pi}_\theta}=e^{U}\pi_\theta\left(\frac{1}{\mu r^2}+\frac{1}{I_{CH_3}}\right),\\
 \dot{\pi}_\theta&=-\frac{\partial \cK}{\partial \theta}=
 -\frac{1}{2}e^{U}\frac{\partial U}{\partial \theta}\left(\frac{\pi_r^2}{\mu}+\pi_\theta^2\left(\frac{1}{\mu r^2}+\frac{1}{I_{CH_3}}\right)\right) -\frac{1}{2}e^{-U}\frac{\partial U}{\partial \theta}.
 \label{eq:chesKHameq}
 \end{split}
\end{equation}

To achieve greater numerical precision, it is preferable to integrate the equations of motion in $(r,p_r,\theta,p_\theta)$ coordinates instead of $(r,\pi_r,\theta,\pi_\theta)$. Equations \eqref{eq:chesKHameq} transform using \eqref{eq:isotransf} to
\begin{equation}
 \begin{split}
  \dot{r}&=\frac{p_r}{\mu},\\
 \dot{p}_r&= p_r\left(\frac{\partial U}{\partial r}\dot{r}+\frac{\partial U}{\partial \theta}\dot{\theta}\right) + \frac{1}{\mu r^3}p_\theta^2 -\frac{\partial U}{\partial r},\\
 \dot{\theta}&=p_\theta\left(\frac{1}{\mu r^2}+\frac{1}{I_{CH_3}}\right),\\
 \dot{p}_\theta&= p_\theta\left(\frac{\partial U}{\partial r}\dot{r}+\frac{\partial U}{\partial \theta}\dot{\theta}\right) -\frac{\partial U}{\partial \theta}, 
 \label{eq:chesKHameqreduced}
 \end{split}
\end{equation}
where we used the isokinetic constraint \eqref{eq:constantT} equivalent to $\cK=0$.

\begin{figure}
 \includegraphics[width=.6\textwidth]{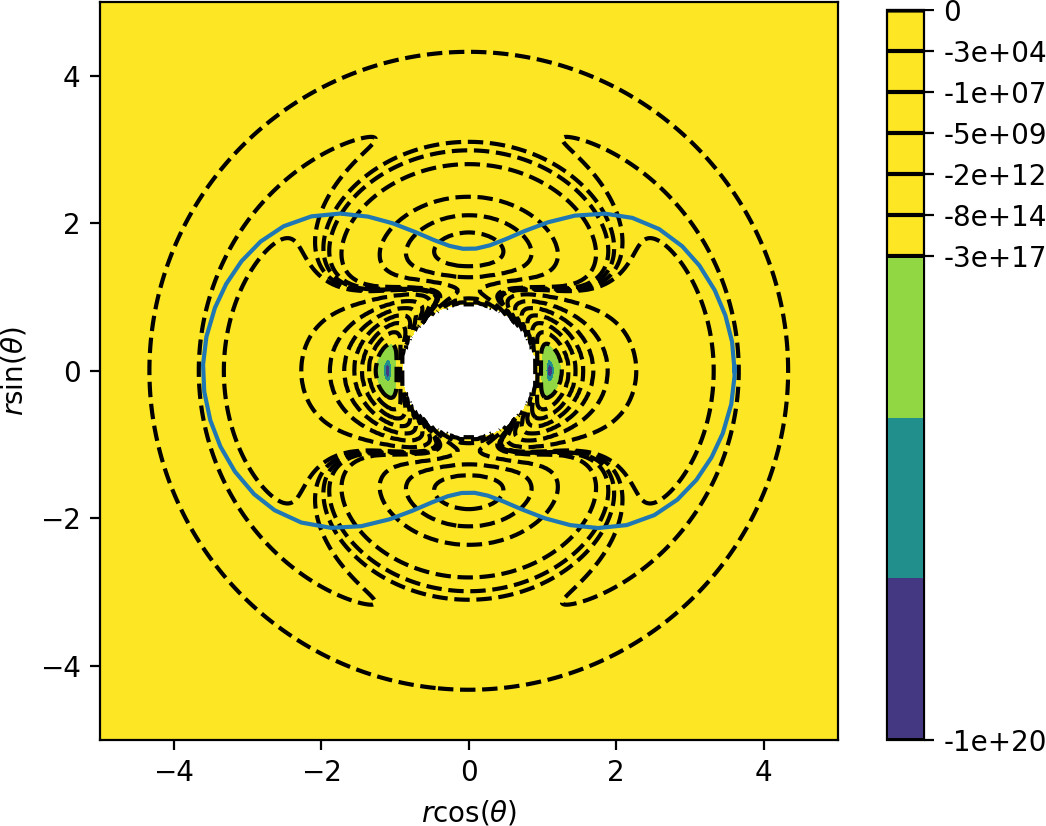}
 \caption{Inner periodic orbit on the potential energy surface $- \frac{1}{2}e^{-U}$.}
 \label{fig:disI}
\end{figure}

The potential $- \frac{1}{2}e^{-U}$ has the same critical points and characteristics as $U$, but the wells are considerably deeper and have steeper walls. In contrast to the microcanonical case, the isokinetic model only possesses two periodic orbits with period $2\pi$ and due to constant nonzero kinetic energy does not admit self-retracing orbits (also referred to as brake orbits) such as $\Gamma^i$ introduced in Section \ref{sec:roamingChes}. One of the periodic orbits delimits the potential wells, see Fig. \ref{fig:disI}; we therefore refer to it as the inner orbit.

The outer orbit, beyond which trajectories do not return and corresponds to the dissociated state of the molecule, is associated with a centrifugal barrier. It is rotationally symmetric and has a radius $r\approx13.43$. Its existence can be proven using a similar argument as in the original system \cite{krajnak2018phase}: suppose $r$ is sufficiently large so that $U$ is effectively independent of $\theta$. Denote $r_{po}$ the solution of
\begin{equation}
\frac{1}{\mu r_{po}^3}p_\theta^2 -\frac{\partial U}{\partial r}=0.
\end{equation}
Then the equations \eqref{eq:chesKHameqreduced} admit a rotationally symmetric periodic orbit with $\dot{\theta}=const$, provided
\begin{equation}
 \begin{split}
  \dot{r}&=0,\\
 \dot{p}_r&=0.
 \end{split}
\end{equation}
This is satisfied by the initial condition $r=r_{po}$, $p_r=0$ and $p_\theta$ given implicitly by $\cK=0$ for any $\theta$. The existence of $r_{po}$ is guaranteed for the potential $U$ and any other potential with leading order term $-cr^{-(2+\varepsilon)}$ for large $r$, with $c>0$ and $\varepsilon>0$.

Both these orbits are unstable, with the largest eigenvalue of the inner orbit under the return map being of the order $10^{21}$. This large instability poses a serious challenge to calculation of its invariant manifolds that guide trajectories in phase space.

\subsection{Classification of trajectories and roaming in the isokinetic Chesnavich model}
\label{subsec:results}

In this section we investigate dynamics in phase space and show the presence of roaming in the thermostatted Chesnavich model. Due to the strength of instability of the inner periodic orbit, we visualise phase space structures using escape times and Lagrangian descriptors\cite{madrid2009ld, lopesino2015cnsns} on surfaces of section rather than calculating invariant manifolds themselves.

\begin{figure}
 \includegraphics[width=0.49\textwidth]{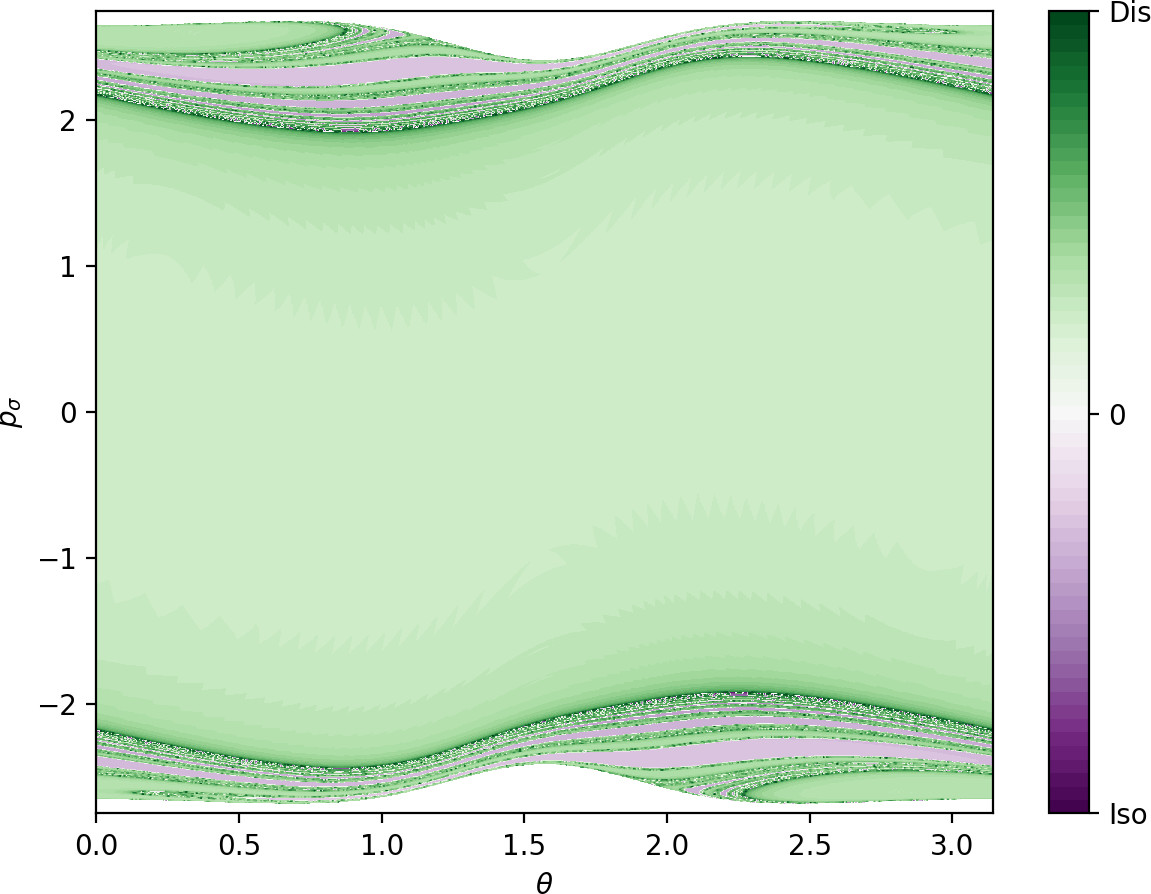}
 \includegraphics[width=0.49\textwidth]{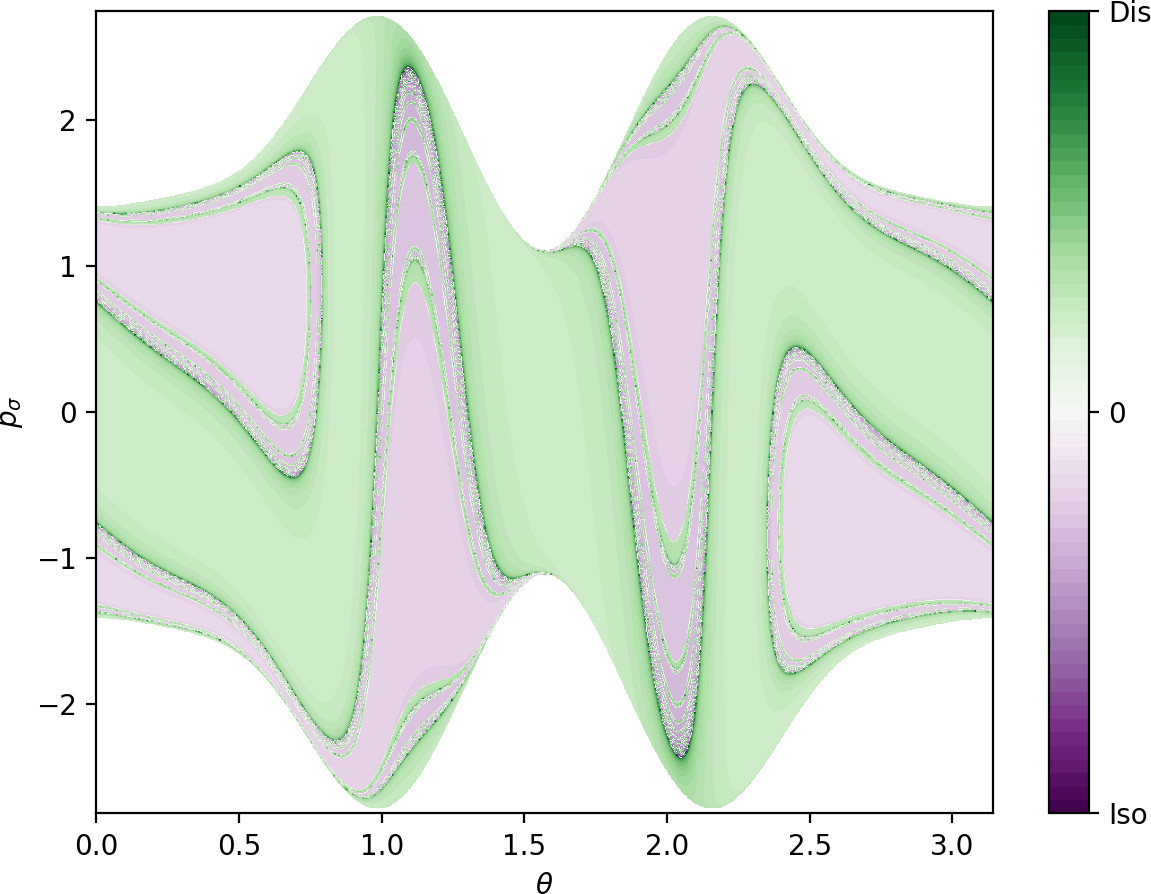}
 \caption{Escape times for initial conditions on the outward annulus of DS$^a$ in Chesnavich's model (left) and on the outward annulus of the inner DS in the isokinetic model (right). Short isomerisation times are shown in light purple, long isomerisation times in dark purple, short dissociation times in light green and long dissociation times in dark green.}
 \label{fig:DSescape}
\end{figure}

The two most natural surfaces of section for this system are: (i) $\theta=0$, $\dot{\theta}>0$ and (ii) the outward annulus of the inner dividing surface (DS).  The latter surface of section is constructed as follows: let the configuration space projection of the inner periodic orbit be parametrised by 
\begin{equation}
  r=\bar{r}(\theta).
  \label{eq:disI}
\end{equation}
Then all points in phase space satisfying \eqref{eq:disI} form a surface with coordinates $\theta$ and the canonically conjugate momentum\cite{Arnold78} $$p_\sigma=p_\theta+\bar{r}'(\theta)p_r.$$
We remark that this surface is, similarly to DS$^a$ and DS$^o$ in Sec. \ref{sec:roamingChes}, a torus. On this torus, the value of $p_r^2$ is given implicitly by $\theta$, $p_\sigma$ and \eqref{eq:constantT} and the sign of $p_r$ is chosen so that
$$\dot{r}>\bar{r}'(\theta),$$
hence the outward direction on the inner DS.

The escape time is defined as the time a trajectory takes to escape the interaction region and reach a dissociated state beyond the outer orbit or either of the wells. Escape time plots obtained for initial conditions on the inner DS in the isokinetic model and on DS$^a$ in the Hamiltonian Chesnavich model are shown in Fig. \ref{fig:DSescape}. The dynamics on DS$^a$ is explained in Sec. \ref{sec:roamingChes}. We can see that while the surfaces share a toroidal geometry, the distributions of escape times of trajectories on them are quite dissimilar. Most of the trajectories in the microcanonical case escape in the same manner as the prototypical dissociation trajectory $\theta=p_\sigma=0$ and more complicated escape dynamics are located near the edges of the surface of section. On the other hand, escape times for the isokinetic model reflect the effects of the constant kinetic energy constraint - note the nearly uniform dissociation around the local maximum at $\theta=\pi/2$ and nearly uniform isomerisation regions around $\theta=1$ and $\theta=2$.

\begin{figure}
 \includegraphics[width=0.49\textwidth]{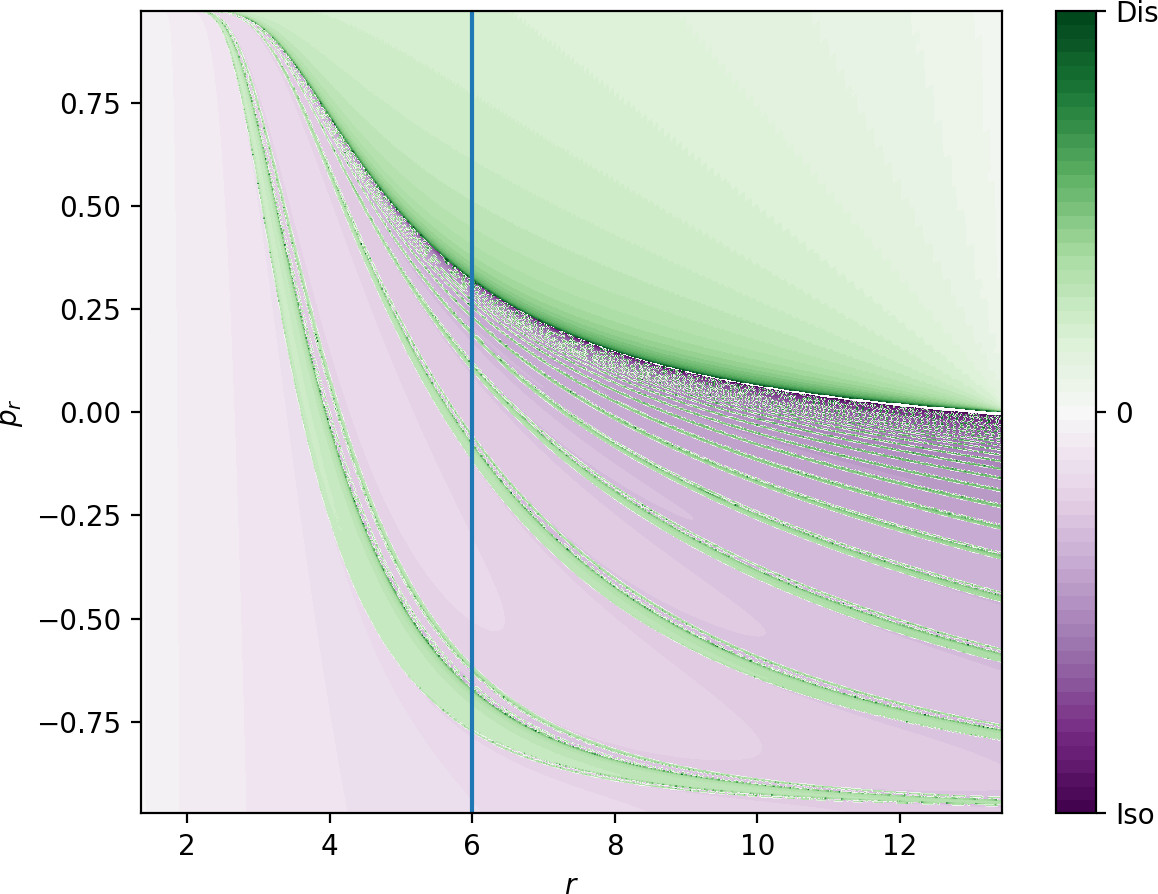}
 \includegraphics[width=0.49\textwidth]{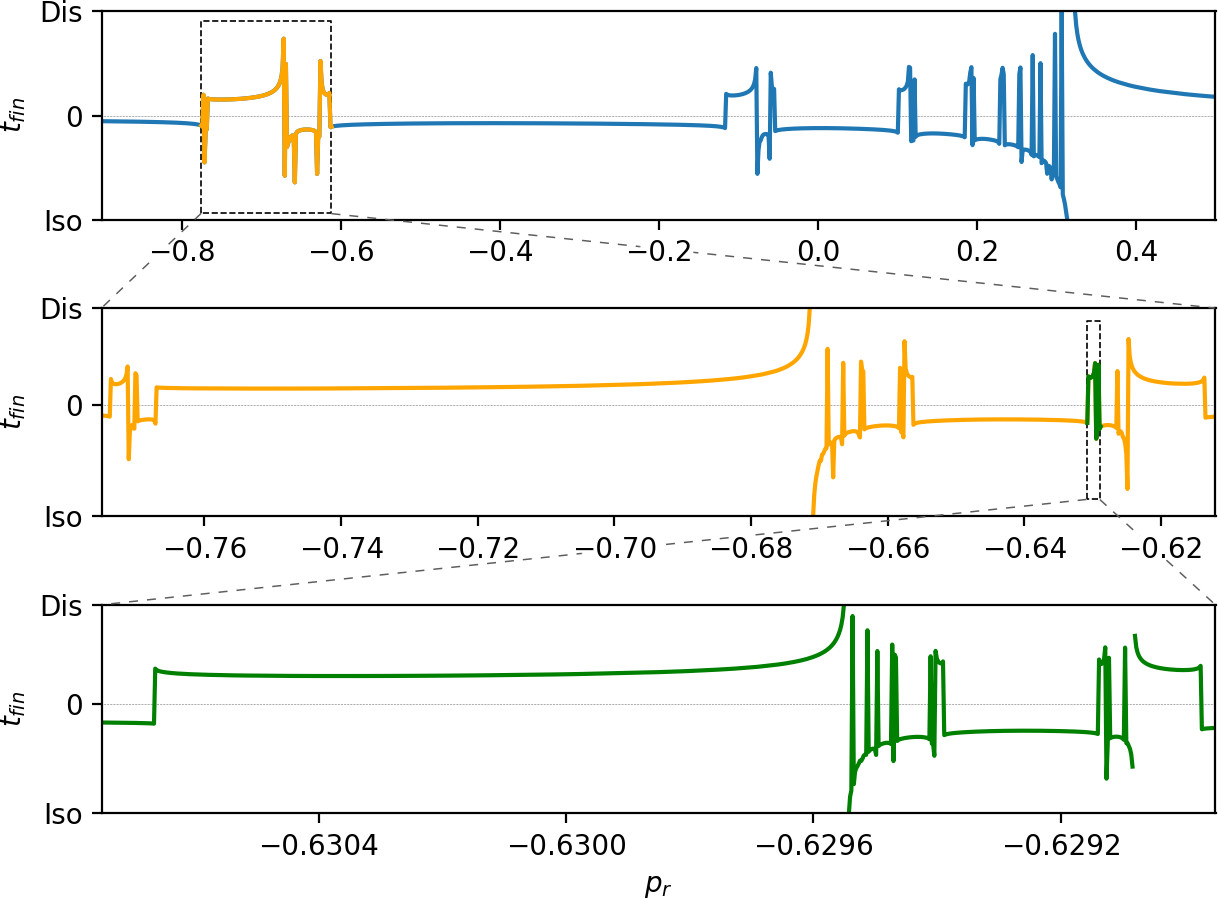}
 \caption{Escape times for initial conditions on $\theta=0$, $\dot{\theta}>0$ and a the profile of the surface for $r=6$ with details of the fractal structure. Short isomerisation times are shown in light purple, long isomerisation times in dark purple, short dissociation times in light green and long dissociation times in dark green.}
 \label{fig:thetaescape}
\end{figure}

The boundary between dissociating and isomerising trajectories, as well as between direct dissociation and more complicated dynamics, consists of invariant manifolds. Fig. \ref{fig:thetaescape} shows escape times on the surface of section $\theta=0$, $\dot{\theta}>0$. Note the singularity in the plot separating a homogeneous area of relatively fast dissociation for large values of $p_r$ from more complicated dynamics for lower values of $p_r$. The singularity can also be seen in the line plot showing escape times on $r=6$, $\theta=0$, $\dot{\theta}>0$.

As it is the case in the microcanonical system, directly dissociating trajectories are guided by the stable invariant manifold of the outer periodic orbit across the phase space bottleneck associated with the outer periodic orbit. Directly dissociating trajectories are located above the stable invariant manifold in Fig. \ref{fig:thetaescape}. All trajectories below the manifold correspond to roaming, isomerizing or nonreactive trajectories. As argued in Sec. \ref{sec:roamingChes}, if there are trajectories on the inner DS that do not dissociate, the invariant manifolds of the inner and outer orbits must intersect. Fig. \ref{fig:DSescape} shows regions of isomerization, therefore the manifolds intersect and all dissociating trajectories in the fractal structures of complicated dynamics on the inner DS correspond to roaming trajectories. 

\begin{figure}
 \includegraphics[width=\textwidth]{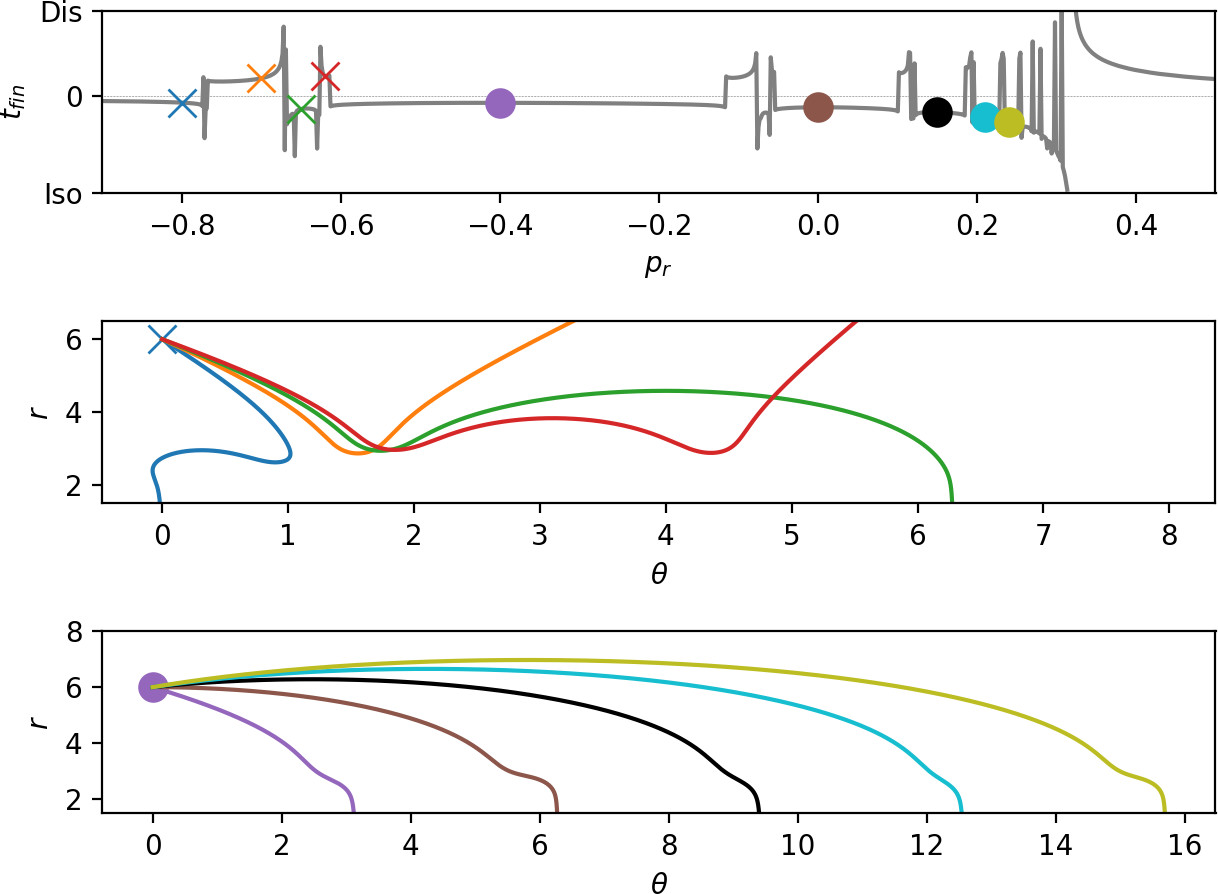}
 \caption{Representative trajectories for different classes along $\theta=0$, $\dot{\theta}>0$, $r=6$.}
 \label{fig:timetraj}
\end{figure}

The bands of isomerizing and dissociating trajectories form a fractal structure and can be classified by the number of rotations in $\theta$ before escaping from the interaction region. These too are separated by invariant manifolds on the inner and outer periodic orbits. The fractal structure from Fig. \ref{fig:thetaescape} can be understood using the configuration space projections of representative trajectories shown in Fig. \ref{fig:timetraj}. In general, trajectories exhibit two dynamical `motifs': they can perform a rotation at large $r$ or interact with the areas of high potential near the local maxima at small $r$. Interaction with areas of high potential leads to escape out of the interaction region, capture in one of the potential wells, rotation in the same direction or rotation in the opposite direction. Trajectories are grouped by the sequence of these motifs in the fractal structure, for example trajectories that rotation by $\pi/2$ and then dissociate or isomerise (see second panel in Fig. \ref{fig:timetraj}) are close to each other and well separated from trajectories that rotate by $\pi$. Each class could be denoted by a sequence of integers denoting the number of rotations (see third panel in Fig. \ref{fig:timetraj}) between interactions with the areas of high potential.

In this way we can find trajectories that perform any possible combination of rotations in the flat area of the interaction region and return any given number of times to the areas of high potential. In other words, regardless of the lack of self-retracing orbits in the interaction region, the scale of complicated dynamics exhibited by the isokinetic system is the same as in the microcanonical system.

\begin{figure}
 \includegraphics[width=0.49\textwidth]{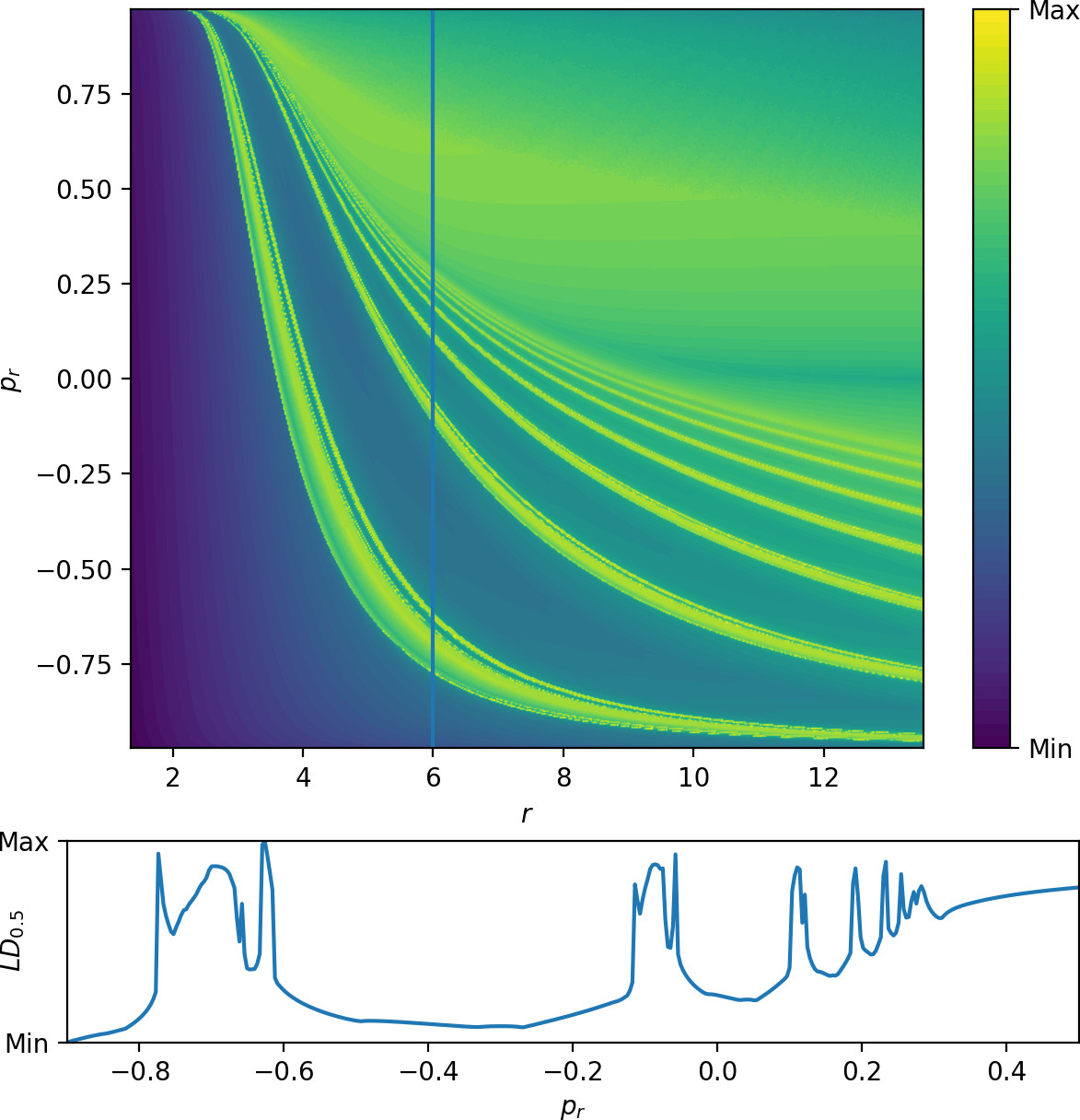}
 \includegraphics[width=0.49\textwidth]{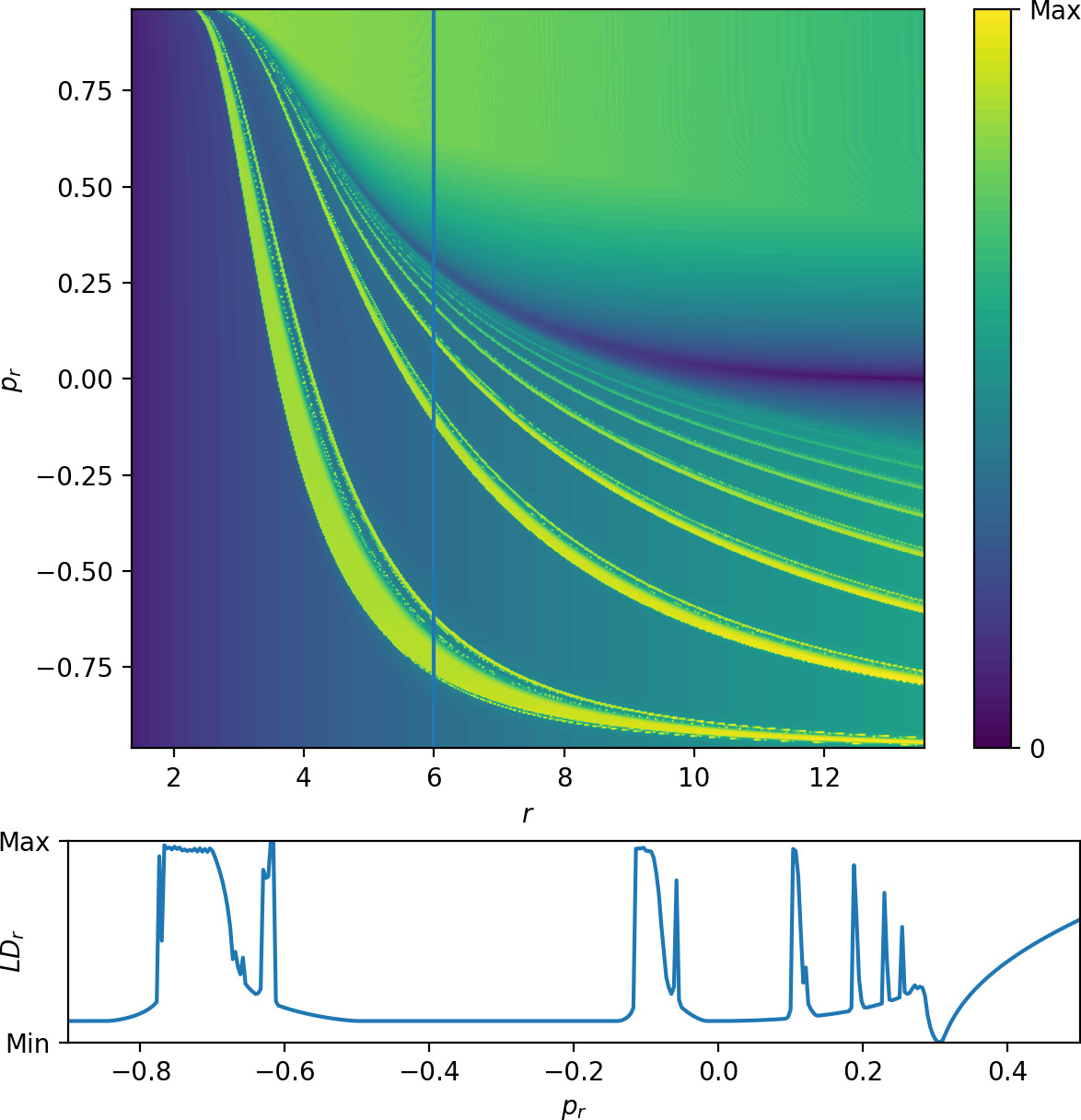}
 \caption{Lagrangian descriptors \eqref{eq:ld05} (left) and \eqref{eq:ldr} (right) for initial conditions on $\theta=0$, $\dot{\theta}>0$ and their profiles for $r=6$.}
 \label{fig:thetald}
\end{figure}

It is important to note that the same results can be obtained using Lagrangian descriptors\cite{madrid2009ld,lopesino2015cnsns} as shown in Fig. \ref{fig:thetald}. We present the plot for the forward Lagrangian descriptor
\begin{equation}
 \int\limits^{40}_{0} \left(|\dot{r}|^{\frac{1}{2}}+|\dot{p_r}|^{\frac{1}{2}}+|\dot{\theta}|^{\frac{1}{2}}+|\dot{p_\theta}|^{\frac{1}{2}}\right) dt,
 \label{eq:ld05}
\end{equation}
with a cut-off at $r=1.1$ for numerical reasons. As opposed to escape times, this Lagrangian descriptor is well defined on invariant manifolds and attains a local minimum on them. Note that the invariant manifolds are visible for an integration time less than six times the period of the inner periodic orbit.
The invariant manifolds are even more pronounced for the radial gain Lagrangian descriptor
\begin{equation}
 \int\limits^{40}_{0} |\dot{r}| dt.
 \label{eq:ldr}
\end{equation}

\section{Conclusions and Outlook}
\label{sec:conc}

In this paper we have studied the nonholonomic dynamics of a Hamiltonian system under the constraint of constant kinetic energy enforced by a Hamiltonian isokinetic thermostat. The thermostatted dynamics, if ergodic, generates a canonical (constant temperature) 
distribution in the system configuration space. Changing the characteristic temperature is equivalent to time scaling.

We further investigated the roaming mechanism in Chesnavich's model for an ion-molecule reaction 
subject to an isokinetic thermostat. Imposing the nonholonomic constraint does not prevent 
the system from exhibiting roaming dynamics, where the origin of the roaming mechanism 
turns out to be analogous to that found in the Hamiltonian case.

The nonexistence of so-called ``brake orbits'' in the isokinetic case (periodic 
orbits with points of zero velocity) leads to differences in
the detailed phase space structure as compared to the microcanonical case,
but the qualitative description of the roaming mechanism as 
a result of trapping in
a region of phase space demarcated by invariant objects remains
unchanged.

%

\section*{FUNDING}
We acknowledge the support of  EPSRC Grant no. EP/P021123/1
and Office of Naval Research (Grant No.~N00014-01-1-0769).

\section*{CONFLICT OF INTEREST}
The authors declare that they have no conflicts of interest.

%
%
%

\end{document}